\documentclass[amsmath,amssymb,11pt,superscriptaddress,reprint, preprintnumbers, notitlepage,aps,twocolumn,nofootinbib]{revtex4-1}
\pdfoutput=1 
\usepackage[utf8]{inputenc}
\usepackage[english]{babel}
\usepackage{amsmath}
\usepackage{graphicx}
\usepackage{dcolumn}
\usepackage{pbox}
\usepackage{amssymb}
\usepackage{epsfig}
\usepackage[dvipsnames]{xcolor}
\usepackage[normalem]{ulem}
\usepackage{slashed}
\usepackage{amssymb}
\usepackage{mathrsfs}
\usepackage{color}
\usepackage[font=small]{caption}
\usepackage[font=small]{subcaption}
\usepackage{url}
\usepackage[makeroom]{cancel}
\usepackage{tikz}
\usetikzlibrary{arrows}
\usepackage{enumitem} 
\usepackage{soul}
\usepackage{ulem}

\definecolor{MyDarkBlue}{rgb}{0.1, 0.1, 0.8} 
\definecolor{SBlue}{rgb}{0.2, 0.4, 0.7} 
\definecolor{MyLightBlue}{rgb}{0.22,0.51,0.9}
\definecolor{MyGreen}{rgb}{0.0, 0.5, 0.0}
\definecolor{BrickRed}{rgb}{0.8, 0.25, 0.33}
\RequirePackage{hyperref}
\hypersetup{colorlinks, citecolor=SBlue,linkcolor=MyDarkBlue, urlcolor=PineGreen}

\usepackage{tikz}
\usepackage{tikz-feynman} 
\tikzfeynmanset{compat=1.1.0}
\makeatletter
\renewcommand\@makecaption[2]{%
  \par
  \vskip\abovecaptionskip
  \begingroup
  
   \small\rmfamily
    \begingroup
     \samepage
     \flushing
     \let\footnote\@footnotemark@gobble
     \@make@capt@title{#1}{#2}\par
    \endgroup
  \endgroup
  \vskip\belowcaptionskip
}
\makeatother

\DeclareUnicodeCharacter{2212}{-}
\begin{document}
\title{\vspace{1cm}\Large 
Weak Triplet Models of Neutrino Magnetic Moments}

\author{\bf Svjetlana Fajfer}
\email[E-mail:]{svjetlana.fajfer@ijs.si}
\affiliation{Jožef Stefan Institute, Jamova 39, P.\ O.\ Box 3000, SI-1001 Ljubljana, Slovenia}
\affiliation{Department of Physics, University of Ljubljana, Jadranska 19, 1000 Ljubljana, Slovenia}

\author{\bf Shaikh Saad}
\email[E-mail:]{shaikh.saad@ijs.si}
\affiliation{Jožef Stefan Institute, Jamova 39, P.\ O.\ Box 3000, SI-1001 Ljubljana, Slovenia}

\begin{abstract}
Experimental limits on neutrino magnetic moments remain several orders of magnitude above the predictions of the Standard Model; therefore, any future detection would provide unambiguous evidence for new physics. In models with Dirac neutrinos, however, mechanisms that enhance the magnetic moment typically generate excessively large neutrino masses. Recently, it has been argued that in frameworks where neutrinos mix with weak-triplet Dirac fermions, the magnetic moment can be decoupled from the neutrino mass. In this work, we revisit this possibility and show that sizable enhancements remain highly nontrivial to realize naturally. We demonstrate that, although the minimal realization allows the magnetic moment to be decoupled from the neutrino mass, obtaining an observable enhancement requires a delicate adjustment of the model parameters. Moreover, in extended scenarios, the decoupling no longer persists: the magnetic moment and neutrino mass become intrinsically linked, such that attempts to enhance the former inevitably induce large contributions to the latter. 
\end{abstract}

\maketitle
\section{Introduction}
The study of neutrino properties has played a central role in uncovering physics beyond the Standard Model (SM). The observation of neutrino oscillations has firmly established that neutrinos are massive, thereby requiring an extension of the SM, which predicts strictly massless neutrinos. Another intriguing property of neutrinos is their magnetic moment~\cite{Pauli:1930pc,Carlson:1932rk,Bethe_1935}. Within the minimally extended SM with  the additional three right-handed neutrinos, to generate  Dirac neutrinos mass terms, the neutrino magnetic moment (NMM) is predicted to be extremely small, of order $\mu_\nu \sim 10^{-19}\mu_B$~\cite{Giunti:2014ixa,Fujikawa:1980yx,Pal:1981rm,Shrock:1982sc,Dvornikov:2003js}, where $\mu_B$ denotes the Bohr magneton. On the other hand, if neutrinos are Majorana, their transition magnetic moments resulting from SM interactions are even smaller  $\mu_\nu \sim 10^{-23}\mu_B$~\cite{Pal:1981rm}. Current experimental limits, however, remain several orders of magnitude above these predictions, with sensitivities reaching $\mu_\nu \lesssim 10^{-11}\mu_B$\footnote{
The current best laboratory bound on the neutrino magnetic moment comes from the GEMMA experiment~\cite{Beda:2012zz}. Astrophysical limits are generally more stringent, reaching $\mu_\nu \lesssim 10^{-12}\mu_B$; however, they are subject to significant astrophysical uncertainties. For a comprehensive compilation of constraints from both laboratory experiments and astrophysical observations, see, for example, Ref.~\cite{Giunti:2024gec}.  Bounds of similar order also arise from cosmological considerations, as discussed, for example, in Ref.~\cite{Li:2022dkc}.
}. This large gap leaves considerable room for the observation of a nonzero neutrino magnetic moment, which would constitute a clear signal of new physics.

From a theoretical perspective, generating a large NMM is straightforward in many extensions of the SM featuring new physics at the TeV scale. However, such enhancements are generally accompanied by a significant drawback: owing to the similar chirality-flipping structure of the neutrino mass and magnetic moment operators, enhancing the latter can induce neutrino masses beyond phenomenologically acceptable limits in the absence of fine-tuning. This tension, commonly referred to as the magnetic moment–mass problem, constitutes a major obstacle to constructing viable models with large neutrino magnetic moments.  If neutrinos are Majorana particles, the NMM operator is flavor antisymmetric, whereas the mass operator is flavor symmetric; this distinction permits~\cite{Bell:2005kz,Davidson:2005cs,Bell:2006wi} the NMM to be enhanced by several orders of magnitude in models of Majorana neutrinos.   In the literature, various mechanisms have been proposed to circumvent this problem and enhance the NMM; see, for example, Ref.~\cite{Voloshin:1987qy,Barbieri:1988fh,Babu:1989wn,   Ecker:1989ph, Babu:1989px, Leurer:1989hx, Babu:1990wv, Chang:1990uga, Choudhury:1990fdr,  Barr:1990um,  Pal:1991qr, Babu:1992vq,  Boyarkin:2014oza,Lindner:2017uvt,Babu:2020ivd}. In this work, we  focus on Dirac neutrinos, for which achieving an enhanced NMM is known to be considerably more challenging~\cite{Bell:2005kz}.

In this study, we explore a framework—the `weak triplet mechanism'—that aims~\cite{Pages:2025odc} to address the longstanding challenge of generating large Dirac neutrino magnetic moments without inducing unacceptably large neutrino masses. In this setup, neutrinos are Dirac particles and therefore require right-handed partners, \(\nu_R\), of the SM neutrinos, \(\nu_L\). Additionally, a pair of weak triplet Dirac fermions, with both left- and right-handed components, is introduced; these states mix  with \(\nu_L\) and \(\nu_R\), respectively. This mixing lies at the heart of the mechanism, enabling new interactions that have been claimed~\cite{Pages:2025odc} to break the conventional proportionality between the neutrino mass and magnetic moment. We perform a careful analysis of this setup and find that electroweak (EW) symmetry breaking—required to generate nonzero neutrino masses—inevitably reintroduces a connection between the neutrino mass and magnetic moment, thereby making it difficult to avoid fine-tuning.

Our findings are summarized as follows: In the minimal realization, the magnetic moment can in principle be decoupled from the neutrino mass; however, achieving a sizable enhancement still requires a tuning of the model parameters. Going beyond the minimal scenario, we find that this decoupling is generically lost: in all extended realizations, the magnetic moment remains tied to the neutrino mass, and its enhancement inevitably reintroduces a corresponding fine-tuning problem.

The rest of this paper is organized as follows. In Sec.~\ref{sec:02}, we present the general framework and analyze the minimal model. After working out the details, we explicitly demonstrate the fine-tuning conditions arising from the mixing between the SM Higgs and the neutral BSM scalar. At the end of this section, we also discuss a scenario in which such mixing is forbidden in an extended model with a discrete \(\mathbb{Z}_2\) symmetry. In Sec.~\ref{sec:03}, we introduce a model with new colored states, in which such mixing with the SM Higgs is automatically forbidden; we then analyze the one-loop diagrams generating both the neutrino mass and magnetic moment, and highlight the issues that undermine the weak triplet mechanism and lead to fine-tuning. Finally, we present our conclusions in Sec.~\ref{sec:04}.

\section{Minimal Model}\label{sec:02}
In this section, we present the general framework of the weak triplet mechanism and analyze its minimal realization. Independently of the details of any specific model, the key ingredients are the introduction of (i) three generations of right-handed neutrinos and (ii) a pair of weak triplet Dirac fermions with left- and right-handed components. In the minimal realization, an additional scalar field in the same representation as the BSM Dirac fermion is also required.

First, we list the quantum numbers of the SM leptons and the SM Higgs under the SM gauge group $SU(3)_c\times SU(2)_L\times U(1)_Y$ and  the global $U(1)_l$ symmetry, where $l$ denotes  the lepton number. Throughout, we assume that lepton number is conserved. 
\begin{align}
&L_L\sim (1,2,-1/2;1)= (\nu_L, \ell_L)^T, 
\\&
\ell_R\sim (1,1,-1;1),
\\&
H\sim (1,2,1/2;0) = (H^+, H^0)^T.
\end{align}
Furthermore, the quantum numbers of the aforementioned BSM states are as follows:
\begin{align}
&\nu_R\sim (1,1,0;1),\\& \Psi_{L,R}\sim (1,3,0;1), 
\\&S\sim (1,3,0;0).
\end{align}
We label this minimal model as Model-I. 
The components of the triplets can be  explicitly written as
\begin{align}
&\Psi=\begin{pmatrix}
\Psi^0/\sqrt{2}  & \Psi^+
\\
\Psi^-  & - \Psi^0/\sqrt{2}
\end{pmatrix}, 
\\&
S= \begin{pmatrix}
S^0/\sqrt{2}  & S^+
\\
S^-  & - S^0/\sqrt{2}
\end{pmatrix}. 
\end{align}
With these particles, the Yukawa part of the Lagrangian consists of the following terms:
\begin{align}
&\mathcal{L}_Y \supset \mathcal{L}_\mathrm{SM} +\mathcal{L}_D + \mathcal{L}_\Psi + \mathcal{L}_S,
\\
&\mathcal{L}_\mathrm{SM}\supset y_e \overline L_L H \ell_R +h.c.,
\\
&\mathcal{L}_D= y_D \overline L_L \widetilde H \nu_R +h.c.,
\\
&\mathcal{L}_\Psi=   m_\Psi \;\mathrm{tr}\left(\overline \Psi_L \Psi_R\right)  + y_\Psi \;\mathrm{tr}\left(\widetilde H \overline L_L  \Psi_R\right) +h.c.  ,
\\
&\mathcal{L}_S=   y_T \;\mathrm{tr}\left(\overline \Psi_L S \Psi_R\right)  + y_\delta \;\mathrm{tr}\left(\overline \Psi_L S\right) \nu_R  +h.c.,
\end{align}
where we have defined $\widetilde H= \epsilon_2 H^*$.

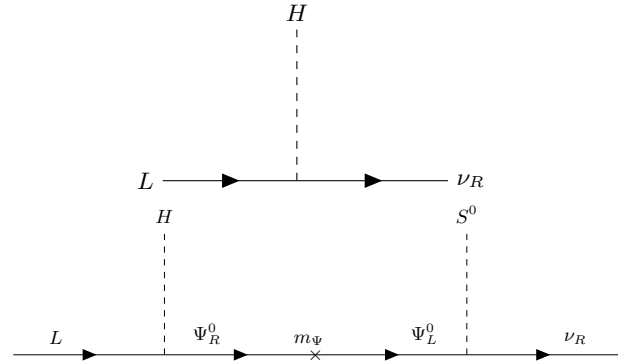
\begin{figure}[h]
\centering
\begin{subfigure}[b]{0.45\textwidth}
  \centering
  \begin{tikzpicture}
    \begin{feynman}
      \vertex (L) {\( L \)};
      \vertex [right=2cm of L] (v);
      \vertex [right=2cm of v] (nuR) {\( \nu_R \)};
      \vertex [above=2cm of v] (H) {\( H \)};

      \diagram* {
        (L) -- [fermion] (v) -- [fermion] (nuR),
        (v) -- [scalar] (H),
      };
    \end{feynman}
  \end{tikzpicture}

\end{subfigure}
\hfill
\begin{subfigure}[b]{0.45\textwidth}
  \centering
  \scalebox{0.8}{
    \begin{tikzpicture}
      \begin{feynman}
        \vertex (L) at (0,0);
        \vertex [right=2.5cm of L] (vL);
        \vertex [right=2.5cm of vL] (v1);
        \vertex [right=2.5cm of v1] (v2);
        \vertex [right=2.5cm of v2] (nuR);
        \vertex [above=2cm of vL] (HL);
        \vertex [above=2cm of v2] (H);

        \node[above=2pt of L, xshift=20pt] {\( L \)};
        \node[above=2pt of vL, xshift=20pt] {\( \Psi_R^0 \)};
        \node[above=2pt of v2, xshift=-20pt] {\( \Psi_L^0 \)};
        \node[above=2pt of nuR, xshift=-20pt] {\( \nu_R \)};
        \node[above=2pt of HL] {\( H \)};
        \node[above=2pt of H] {\( S^0 \)};
        \node at (v1) {\( \times \)};
        \node[above right=2pt of v1, xshift=-15pt] {\( m_\Psi \)};

        \diagram* {
          (L) -- [fermion] (vL),
          (vL) -- [scalar] (HL),
          (vL) -- [fermion] (v1),
          (v1) -- [fermion] (v2),
          (v2) -- [fermion] (nuR),
          (v2) -- [scalar] (H),
        };
      \end{feynman}
    \end{tikzpicture}
  }
\end{subfigure}
\caption{Model-I. Tree-level Dirac neutrino mass generation mechanisms: (upper panel) direct coupling via \(y_D\); (bottom panel) contribution mediated by vector-like fermions with a mass insertion \(m_\Psi\).}
\label{fig:tree}
\end{figure}

Moreover,  the scalar potential is given by
\begin{align}
    V&\supset -\mu^2_H H^\dagger H + \lambda_H  \left(H^\dagger H\right)^2   -\mu^2_S \;\mathrm{tr}\left(S^2\right) + \lambda_S \;\mathrm{tr}\left(S^2\right)^2  \nonumber
 \\&
     + \mu H^\dagger S H  + \lambda_{HS} \left(H^\dagger H\right) \;\mathrm{tr}\left(S^2\right) . \label{eq:fullpot}
\end{align}
Note that there is no symmetry that forbids the cubic term, $\mu$, and its presence leads to an induced VEV (vacuum expectation value) of the scalar $S$. We define $\langle H^0\rangle = v_H\simeq 174$ GeV and $\langle S^0\rangle = v_S$.  In this model, due to $v_S\neq 0$, neutrinos receive an additional tree-level contribution to their mass, beyond the usual Dirac mass term (upper panel of Fig.~\ref{fig:tree}), as shown in the bottom panel of Fig.~\ref{fig:tree}. Consequently, as we will show below, the value of $v_S$ is expected to be quite small. 

The complete mass matrix in the neutral sector at the tree-level, is therefore, given by 
\begin{align}
&M_N= \begin{pmatrix} 
m_D& m_H
\\
m_S&m_\Psi
\end{pmatrix} = \begin{pmatrix} 
y_D v_H& \frac{y_\Psi}{\sqrt{2}}  v_H
\\
y_\delta v_S&m_\Psi
\end{pmatrix} = N_L M_N^\mathrm{diag} N_R^\dagger .
\end{align}
For simplicity, in this work, we consider a single generation of neutrino, an extension to three generations can be trivially obtained. 
For phenomenological reason, we are interested in a region of the parameter space, which corresponds to  $m_\Psi > m_H \gg m_D > m_S$, to be discussed below. Therefore, with $m_H/m_\Psi <1$,   it is straightforward to diagonalize the above matrix to get 
\begin{align}
    M_N^\mathrm{diag}\simeq \begin{pmatrix} 
y_D v_H- \frac{v_H v_S y_\delta y_\Psi}{\sqrt{2} m_\Psi} - \frac{v_H^3 y_D y_\Psi^2}{4 m_\Psi^2}   &0
\\
0&   m_\Psi +  \frac{y^2_\Psi v_H^2}{4 m_\Psi}  
\end{pmatrix}. \label{eq:tree-mass}
\end{align}
Here, the (22)-entry ($\approx m_\Psi$) corresponds to the mass of the heavy BSM neutral state. Throughout the text, we take a benchmark value of $m_\Psi=1$ TeV for this mass unless otherwise specified.
If the first term in the (11)-entry provides the main contribution to the neutrino mass, then  $m_\nu^\mathrm{tree}\simeq y_D v_H$, which leads to   $y_D\sim 3\times 10^{-13}$ to obtain the correct neutrino mass scale of $m_\nu\sim 0.05$ eV. Now, focusing only on the second term  for the light neutrinos, it is important to put the constraint on the VEV of the BSM state $v_S < (\sqrt{2} m_\Psi m_\nu)/(v_H y_\delta y_\Psi)$ such that it does not overshoot neutrino mass scale. By taking benchmark values   $y_\delta y_\Psi\sim 1$ and $y_\delta y_\Psi\sim 10^{-3}$, one obtains $v_S< 4\times 10^{-10}$ GeV and $v_S< 4\times 10^{-7}$ GeV, respectively. The third term, on the other hand, is  negligible.

Furthermore, the flavor eigenstates can be expressed as a linear combination of the mass eigenstates (represented with a hat on top)
\begin{align}
&\begin{pmatrix} 
\nu_{L,R}
\\
\Psi_{L,R}^0
\end{pmatrix}=  N_{L,R} \begin{pmatrix} 
\hat \nu_{L,R}
\\
\hat \Psi_{L,R}^0
\end{pmatrix}, 
\end{align}
with
\begin{align}
&
N_{L,R}= \begin{pmatrix}
        \cos\theta_{L,R}&\sin\theta_{L,R}
        \\
        -\sin\theta_{L,R}&\cos\theta_{L,R}
    \end{pmatrix},
\\    
&\sin\theta_L\simeq    \left( 1-\frac{v_H^2 y_{\Psi }^2}{4 m_{\Psi }^2}  \right) \frac{v_H y_{\Psi }}{   \sqrt{2} m_{\Psi }} \simeq  \frac{v_H y_{\Psi }}{   \sqrt{2} m_{\Psi }},
\\
&\sin\theta_R\simeq \left( 1 - \frac{y_\Psi^2 v_H^2}{2m^2_\Psi}\right) \left( \frac{y_\delta v_S}{m_\Psi}  + \frac{y_Dy_\Psi v_H^2}{\sqrt{2}m^2_\Psi}  \right).
\end{align}

As for the charged leptons, we get
\begin{align}
&M_E= \begin{pmatrix} 
m_E&\sqrt{2} m_H
\\
0&m_\Psi
\end{pmatrix} = E_L M_E^\mathrm{diag} E_R^\dagger ,
\end{align}
with $m_E=y_ev_H$. To the leading order in $m_H/m_\Psi <1$, the masses are given by 
\begin{align}
    m_E^\mathrm{light}\simeq y_e v_H,\quad m_E^\mathrm{heavy}\simeq m_\Psi .
\end{align}
Similarly, the flavor eigenstates can be expressed in terms of the mass eigenstates 
\begin{align}
&\begin{pmatrix} 
\ell_{L,R}^-
\\
\Psi_{L,R}^-
\end{pmatrix}=  E_{L,R} \begin{pmatrix} 
\hat \ell_{L,R}^-
\\
\hat \Psi_{L,R}^-
\end{pmatrix} ,
\end{align}
with
\begin{align}
&E_{L,R}= \begin{pmatrix}
        \cos\theta_{L,R}^E&\sin\theta_{L,R}^E
        \\
        -\sin\theta_{L,R}^E&\cos\theta_{L,R}^E
    \end{pmatrix},
    \\
&\sin\theta_L^E \simeq \frac{v_H y_{\Psi }}{m_{\Psi }}, \;\;
\\
&\sin\theta_R^E \simeq \frac{y_ey_{\Psi }v_H^2 }{m^2_{\Psi }}.
\end{align}

Now, concentrating on the scalar sector, the mass squared matrix for the neutral states,  in the basis $(h^0, S^0)$, takes the following form
\begin{align}
M^2_0= \begin{pmatrix}
    4 v_H^2 \lambda_H & 2\sqrt{2} v_H v_S \lambda_{HS} - v_H \mu
    \\
    2\sqrt{2} v_H v_S \lambda_{HS} - v_H \mu & 8v_S^2 \lambda_S + \frac{v_H^2 \mu}{\sqrt{2}v_S}
\end{pmatrix}. \label{eq:neutral}
\end{align}
In the limit of no mixing between these states, SM Higgs mass is given by
\begin{align}
&    M_h= 2 v_H \lambda_H^{1/2} \simeq 125 \; \mathrm{GeV}.
\end{align}
In general, the physical SM Higgs is identified with the $\phi^0_1$ state, such that 
\begin{align}
&h^0= \cos\zeta \phi^0_1  - \sin\zeta \phi^0_2,
\\
&S^0= \sin\zeta \phi^0_1  + \cos\zeta \phi^0_2,
\end{align}
where,
\begin{align}
  \tan 2 \zeta= \frac{4\sqrt{2} v_H v_S \lambda_{HS} -2 v_H \mu}{8v_S^2 \lambda_S + \frac{v_H^2 \mu}{\sqrt{2}v_S} - 4 v_H^2 \lambda_H }.  \label{eq:mixing-angle}
\end{align}
On the other hand, for the  charged scalar states, we have 
\begin{align}
M^2_{\pm}= \begin{pmatrix}
    \sqrt{2} v_S \mu & v_H \mu
    \\
    v_H \mu & \frac{v_H^2 \mu}{\sqrt{2} v_S}
\end{pmatrix},
\end{align}
and  in the mass basis we introduce $G^\pm, \phi^\pm$
\begin{align}
&H^\pm= \cos\xi G^\pm  - \sin\xi \phi^\pm,
\\
&S^\pm= \sin\xi G^\pm  + \cos\xi \phi^\pm,
\end{align}
with
\begin{align}
  \tan 2 \xi= \frac{2\sqrt{2}v_H v_S}{v_H^2-2v_S^2}.
\end{align}
The Goldstone, $G^\pm$, is massless and the physical singly charged state, $\phi^\pm$, has the following mass
\begin{align}
    M^2_{\phi^{\pm}} = \frac{v_H^2 \mu}{\sqrt{2}v_S} + \sqrt{2} v_S \mu . \label{eq:charged}
\end{align}
For a benchmark value  $M_{\phi^{\pm}}= 1$ TeV, and using the above upper bounds   $v_S\sim 4\times 10^{-10}$ GeV  and $v_S\sim 4\times 10^{-7}$ GeV, we obtain  $\mu \sim 1.9\times 10^{-8}$ GeV and $\mu \sim 1.9\times 10^{-5}$ GeV, respectively. Therefore, comparing Eq.~\eqref{eq:charged} and Eq.~\eqref{eq:neutral}, it is evident that $\phi^{\pm}$ and $\phi^{0}$ are practically degenerate in mass. Consequently, the contributions to the oblique parameters~\cite{Peskin:1990zt,Peskin:1991sw} are suppressed and do not impose meaningful constraints on the model parameters.

With these ingredients, we now present the relevant part of the Lagrangian in the mass basis required for the computation of the neutrino mass and NMM. We keep these terms to the leading order in $\sin\theta_X$ ($X=L,R$).
\begin{widetext}
\begingroup
\allowdisplaybreaks
\begin{align}
m_\Psi \overline \Psi_L \Psi_R      \label{eq:100}
&\supset
m_\Psi  \overline{\hat\Psi^0}_L \hat\Psi^0_R +  m_\Psi \overline{\hat \Psi^\pm}_L \hat\Psi^\pm_R 
\\   
y_D \overline L_L \widetilde H \nu_R 
&\supset   
m_D  \overline{\hat \nu}_L \hat\nu_R 
+
\frac{y_D }{\sqrt{2}} \left(  \cos\zeta \phi_1  - \sin\zeta \phi_2  \right) \bigg\{   
\overline{\hat \nu}_L \hat\nu_R
+
\sin\theta_R \overline{\hat \nu}_L \hat\Psi^0_R 
+   
\sin\theta_L    \overline{\hat \Psi^0}_L \hat\nu_R 
\bigg\}
\\   
y_\Psi \overline L_L \widetilde H \Psi_R  
&\supset   
\frac{y_\Psi}{2}  \left(  \cos\zeta \phi_1  - \sin\zeta \phi_2  \right) 
\bigg\{
\overline{\hat \nu}_L \hat\Psi^0_R
-
\sin\theta_R  \overline{\hat \nu}_L \hat\nu_R 
+
\sin\theta_L \overline{\hat \Psi^0}_L \hat\Psi^0_R
-   
\sin\theta_L
\sin\theta_R
\overline{\hat \Psi^0}_L \hat\nu_R
\bigg\}
\\   
y_T \overline \Psi_L S \Psi_R   
&\supset 
\frac{y_T}{\sqrt{2}} \cos\xi  \left( \overline{\hat \Psi^-}_L  \phi^- - \overline{\hat \Psi^+}_L \phi^+ \right) \hat\Psi^0_R
+
\frac{1}{\sqrt{2}}y_T \cos\xi 
\bigg\{
\sin\theta_L   \overline{\hat\nu}_L \left(\hat \Psi^+_R  \phi^- - \hat \Psi^-_R \phi^+ \right)  \nonumber
\\&+ 
\sin\theta_R  \left(- \overline{\hat \Psi^-}_L  \phi^- + \overline{\hat \Psi^+}_L \phi^+ \right) \hat\nu_R 
\bigg\}
\\   
y_\delta \overline \Psi_L S \nu_R   
&\supset 
y_\delta \cos\xi
\sin\theta_R \left( \overline{\hat \Psi^-}_L  \phi^- + \overline{\hat \Psi^+}_L \phi^+ \right) \hat\Psi^0_R
 -y_\delta   
\sin\theta_L
\overline{\hat\nu}_L
\bigg\{
\hat\nu_R +   \sin\theta_R   \hat\Psi^0_R
\bigg\} \left(  \sin\zeta \phi_1  + \cos\zeta \phi_2  \right) \nonumber
\\&
+ y_\delta
\bigg\{
\left( \overline{\hat \Psi^-}_L  \phi^- + \overline{\hat \Psi^+}_L \phi^+ \right) \cos\xi
+ 
\left( \overline{\hat \Psi^0}_L  -  \sin\theta_L   \overline{\hat\nu}_L \right) \left( \sin\zeta \phi_1  + \cos\zeta \phi_2 \right) 
\bigg\}
\hat\nu_R  \label{eq:101}
\end{align}
\endgroup

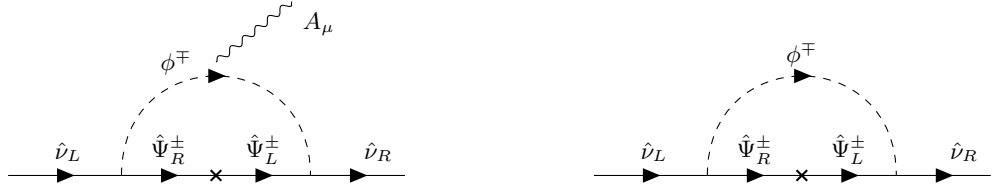
\begin{figure}[h]
\centering
\begin{tikzpicture}
  \begin{feynman}
    \vertex (a1) at (0,0);                          
    \vertex [right=1.5cm of a1] (aPsi);             
    \vertex [right=2.5cm of aPsi] (aF);             
    \vertex [right=1.25cm of aF] (anuR);            

    \coordinate (amidferm) at ($(aPsi)!0.5!(aF)$);   

    \vertex (aSloopTop) at ($(aPsi)!0.5!(aF)+(0,1.5)$);
    \vertex (aboson) at ($(aSloopTop)+(1,0.8)$);    

    \draw[thick] ([xshift=-2pt,yshift=-2pt]amidferm) -- ([xshift=2pt,yshift=2pt]amidferm);
    \draw[thick] ([xshift=-2pt,yshift=2pt]amidferm) -- ([xshift=2pt,yshift=-2pt]amidferm);

    \node[above=2pt of aPsi, xshift=-20pt] {\(\hat\nu_L\)};
    \node[above=2pt of anuR, xshift=-10pt] {\(\hat\nu_R\)};
    \node[above=-8pt of aSloopTop, xshift=-15pt] {\(\phi^\mp\)};
    \node[below right=1pt and 1pt of aboson] {\(A_\mu\)};
    \node[above=1pt of $(aPsi)!0.5!(amidferm)$] {\(\hat\Psi_R^\pm\)};
    \node[above=1pt of $(amidferm)!0.5!(aF)$] {\(\hat\Psi_L^\pm\)};

    \diagram* {
      (a1) -- [fermion] (aPsi),
      (aPsi) -- [fermion] (amidferm),
      (amidferm) -- [fermion] (aF) -- [fermion] (anuR),
      (aPsi) -- [half left, charged scalar, dashed, looseness=1.8] (aF),
      (aSloopTop) -- [boson] (aboson),
    };

    \vertex (b1) at ($(anuR)+(2.5,0)$);                           
    \vertex [right=1.5cm of b1] (bPsi);             
    \vertex [right=2.5cm of bPsi] (bF);             
    \vertex [right=1.25cm of bF] (bnuR);            

    \coordinate (bmidferm) at ($(bPsi)!0.5!(bF)$);   

    \vertex (bSloopTop) at ($(bPsi)!0.5!(bF)+(0,1.5)$);

    \draw[thick] ([xshift=-2pt,yshift=-2pt]bmidferm) -- ([xshift=2pt,yshift=2pt]bmidferm);
    \draw[thick] ([xshift=-2pt,yshift=2pt]bmidferm) -- ([xshift=2pt,yshift=-2pt]bmidferm);

    \node[above=2pt of bPsi, xshift=-20pt] {\(\hat\nu_L\)};
    \node[above=2pt of bnuR, xshift=-10pt] {\(\hat\nu_R\)};
    \node[above=-5pt of bSloopTop] {\(\phi^\mp\)};
    \node[above=1pt of $(bPsi)!0.5!(bmidferm)$] {\(\hat\Psi_R^\pm\)};
    \node[above=1pt of $(bmidferm)!0.5!(bF)$] {\(\hat\Psi_L^\pm\)};

    \diagram* {
      (b1) -- [fermion] (bPsi),
      (bPsi) -- [fermion] (bmidferm),
      (bmidferm) -- [fermion] (bF) -- [fermion] (bnuR),
      (bPsi) -- [half left, charged scalar, dashed, looseness=1.8] (bF),
    };

  \end{feynman}
\end{tikzpicture}
\caption{Model-I. Left panel: Leading contribution to NMM. The outgoing photon can be emitted from the internal fermion or boson line. Right panel: Vanishing contribution to neutrino mass.}
\label{fig:04}
\end{figure}
\end{widetext}

Given this set of Lagrangian terms Eqs.~\eqref{eq:100}-\eqref{eq:101}, the leading contribution to the NMM is depicted in Fig.~\ref{fig:04} (left panel). Subleading contributions are proportional to the neutrino mass and are therefore omitted. The same diagram, with the external photon leg removed, would naively contribute to the neutrino mass; see Fig.~\ref{fig:04} (right panel). However, it is important to note that the right vertex, proportional to $y_\delta$, yields contributions with the same sign for both diagrams involving $\hat{\Psi}^+$ and $\hat{\Psi}^- $, whereas the left vertex, proportional to $y_\Psi$, gives opposite signs. As a result, the contributions cancel, and the diagram in the right panel does not generate a neutrino mass. In contrast, inserting the external photon legs (as in the NMM diagram) introduces additional sign factors depending on the charges of the particles running in the loop, leading to a nonzero NMM and thereby breaking the proportionality between the neutrino mass and magnetic moment.

\begin{figure}[th!]
\centering
\includegraphics[width=0.35\textwidth]{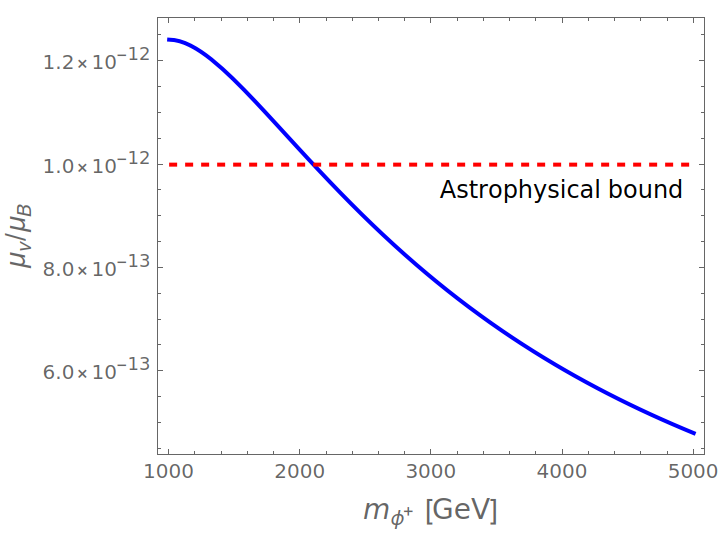}
\caption{Model-I. Magnitude of NMM as a function of the BSM scalar mass (we have fixed $m_\Psi= 1$ TeV).} \label{fig:minimal}
\end{figure}

These diagrams contribute to NMM as~\cite{Lavoura:2003xp} 
\begin{align}
\mu_\nu &\simeq  \frac{y_\Psi\, y_T\, y_\delta}{8\pi^2} v_H \cos\xi \, m^2_{\phi^\pm} \nonumber 
\\&
\frac{  2(m^2_{\phi^\pm}-m^2_\Psi) + (m^2_{\phi^\pm}+m^2_\Psi)  \log \frac{m^2_\Psi}{m^2_{\phi^\pm}} }{(m^2_{\phi^\pm}-m^2_\Psi)^3},
\end{align}
i.e.,
\begin{align}
&\mu_\nu  
\sim \frac{y_\Psi\, y_T\, y_\delta}{8\pi^2} \frac{v_H}{\Lambda^2},
\\
&\delta m^{\Psi^\pm}_\nu =0,
\end{align}
where $\Lambda$ represents the new physics scale.  The above equation shows that the NMM depends on three Yukawa couplings, namely, $\mu_\nu \propto y_\Psi\, y_T\, y_\delta$. In Fig.~\ref{fig:minimal}, we show the magnitude of this NMM by varying the mass of the corresponding scalar field.  

\begin{figure}[th]
\centering
\begin{tikzpicture}
  \begin{feynman}
    \vertex (b1) at (0,0);                           
    \vertex [right=1.5cm of b1] (bPsi);             
    \vertex [right=2.5cm of bPsi] (bF);             
    \vertex [right=1.25cm of bF] (bnuR);            

    \coordinate (bmidferm) at ($(bPsi)!0.5!(bF)$);   

    \vertex (bSloopTop) at ($(bPsi)!0.5!(bF)+(0,1.5)$);

    \draw[thick] ([xshift=-2pt,yshift=-2pt]bmidferm) -- ([xshift=2pt,yshift=2pt]bmidferm);
    \draw[thick] ([xshift=-2pt,yshift=2pt]bmidferm) -- ([xshift=2pt,yshift=-2pt]bmidferm);

    \node[above=2pt of bPsi, xshift=-20pt] {\(\hat\nu_L\)};
    \node[above=2pt of bnuR, xshift=-10pt] {\(\hat\nu_R\)};
    \node[above=-5pt of bSloopTop] {\(\phi_{1,2}^0\)};
    \node[above=1pt of $(bPsi)!0.5!(bmidferm)$] {\(\hat\Psi_R^0\)};
    \node[above=1pt of $(bmidferm)!0.5!(bF)$] {\(\hat\Psi_L^0\)};

    \diagram* {
      (b1) -- [fermion] (bPsi),
      (bPsi) -- [fermion] (bmidferm),
      (bmidferm) -- [fermion] (bF) -- [fermion] (bnuR),
      (bPsi) -- [half left, scalar, dashed, looseness=1.8] (bF),
    };

  \end{feynman}
\end{tikzpicture}
\caption{One-loop neutrino mass in the minimal realization (Model-I).}
\label{fig:009}
\end{figure}
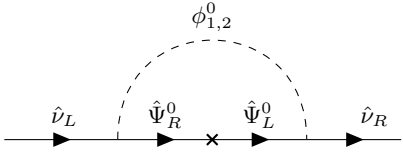

\begin{figure}[th!]
\centering
\includegraphics[width=0.45\textwidth]{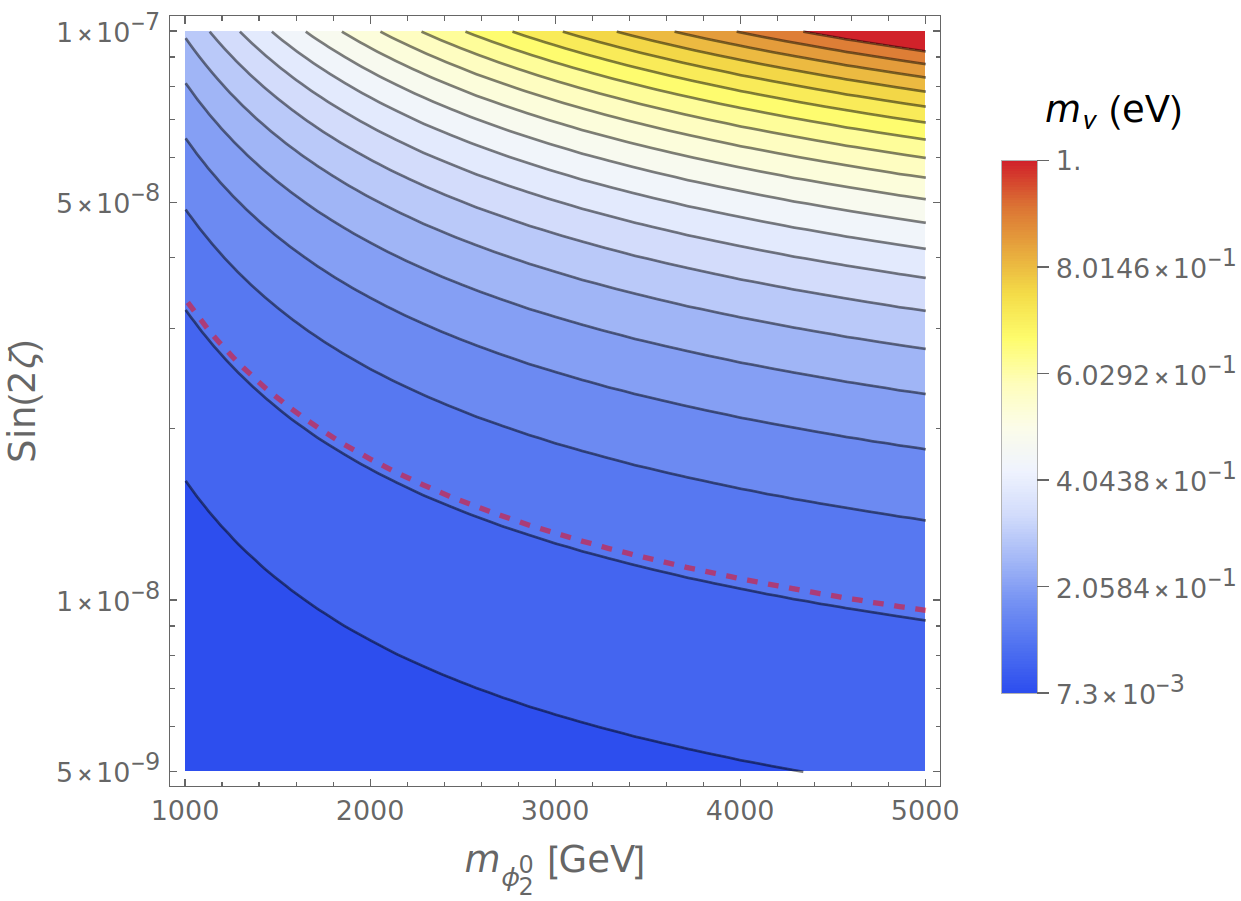}
\caption{Model-I. Fine-tuned scalar mixing parameter as a function of the BSM scalar ($\phi^0_2$) mass, with $m_\Psi = 1$~TeV fixed. The color bar indicates the neutrino mass scale, Eq.~\eqref{eq:numass-minimal}. The red dashed line corresponds to the observed neutrino mass scale $m_\nu = 0.05$ eV. }\label{fig:minimal-mixing}
\end{figure}

It is important to note that, although the diagram in Fig.~\ref{fig:04} (right panel) does not generate a neutrino mass, the mixing between the SM Higgs and the neutral BSM state induces a one-loop contribution to the neutrino mass, through the propagation of scalar $\phi^{0}_{1,2}$ states, as shown in Fig.~\ref{fig:009}. This additional one-loop contribution to the neutrino mass was not taken into account in Ref.~\cite{Pages:2025odc}. The resulting neutrino mass depends on the same unsuppressed Yukawa couplings (required to enhance the NMM), $m_\nu \propto y_\Psi\, y_\delta$, and the same mass scales, $m_\Psi$ and $m_\phi$. However, an additional parameter enters, namely the mixing angle of the neutral states. More explicitly, this contribution is given by
\begin{align}
\delta m^{\Psi^0}_\nu=     \frac{y_\Psi\, y_\delta \sin2\zeta \, m_\Psi}{64\pi^2}   \left(
\frac{m^2_{\phi^0_1} \log \frac{m^2_{\phi^0_1} }{m^2_\Psi} }{m^2_{\phi^0_1}-m^2_\Psi}
-
\frac{m^2_{\phi^0_2} \log \frac{m^2_{\phi^0_2} }{m^2_\Psi} }{m^2_{\phi^0_2}-m^2_\Psi}
\right), \label{eq:numass-minimal}
\end{align}
which highly constraints the relevant model parameters. For $m_\Psi, m_\phi= 1$ TeV, the contribution from the diagram with the SM Higgs running in the loop can be neglected, and the term inside the parentheses reduces to unity.   Assuming, for example,  $y_\Psi, y_\delta\sim \mathcal{O}(1)$ to enhance NMM, one finds that $\sin\zeta \lesssim 10^{-11}$ is required to avoid overshooting the observed neutrino mass scale, namely, $\delta m_\nu < 0.05$ eV. Utilizing the mixing angle defined in Eq.~\eqref{eq:mixing-angle}, this leads to the following  fine-tuning condition on the relevant model parameters:
\begin{align}
2\sqrt{2} v_S\, \lambda_{HS} - \mu \lesssim 5\times 10^{-8}\,{\text GeV}  \label{eq:EWcondition}
\end{align}
By varying the mass of $\phi_2^0$ while fixing $m_\psi = 1$~TeV, Fig.~\ref{fig:minimal-mixing} depicts the degree of fine-tuning required for the mixing parameter $\sin(2\zeta)$.
If this condition is not imposed, then the corresponding enormous contribution must be canceled against the already present tree-level neutrino mass, Eq.~\eqref{eq:tree-mass}, in order to reproduce the observed neutrino mass scale.

We comment that in the limit $\mu \to 0$ (in particular, $\mu \ll 1\,\text{GeV}$) advocated by Ref.~\cite{Pages:2025odc}, Eq.~\eqref{eq:EWcondition} shows that observability of the NMM -- which requires new-physics states around the TeV scale -- instead requires an equally small combination $v_S\lambda_{HS}$. Such smallness does not automatically follow from the theory, but instead points to a very special choice of the relevant parameters. It is important to clarify that the global symmetry argument that may lead to a small $\mu$ has absolutely no impact on the parameter $\lambda_{HS}$.   On the other hand, if $-\mu^2_S>0$   is assumed in Eq.~\ref{eq:fullpot} so that $S$ does not participate in spontaneous symmetry breaking, then only the $\mu$ term survives in Eq.~\ref{eq:EWcondition}. Consequently, the requirement can be satisfied in the limit $\mu\to 0$.

\vspace{10pt}
\textbf{Extended models with color neutral states}-- 
Before concluding this section, it is worth emphasizing that the non-minimal models considered in Ref.~\cite{Pages:2025odc} all contain electrically neutral components that inevitably develop vacuum expectation values (VEVs) and mix with the SM Higgs field. This, in turn, induces additional loop contributions to the neutrino mass via the mixed neutral states, thereby requiring fine-tuning, as discussed above. On the other hand, the last model listed in their Table~I includes an additional $\mathbb{Z}_2$ symmetry, which prevents the neutral component of the inert doublet, $\eta \sim (1,2,1/2;-)$, from mixing with the SM Higgs (here, the minus sign denotes a $\mathbb{Z}_2$-odd state). In addition to the Dirac states $\Psi_{L,R}$, this model also contains another pair of BSM Dirac fermions, $F_{L,R} \sim (1,2,1/2;-)$, transforming as weak doublets. We label this representative model as Model-II.

Let us briefly analyze this model. The relevant part of the Yukawa Lagrangian reads
\begin{align}
\mathcal{L} &\supset y_1 \overline  F_R \eta \Psi_L \supset y_1 \left( \overline{F^+_R}  \eta^+   
- \frac{1}{\sqrt{2}}
 \overline{F^0_R} \eta^0 \right) \Psi^0_L, \nonumber
 \\&
 \supset   - \frac{v_H \,y_{\Psi }\,y_1}{   \sqrt{2} m_{\Psi }}   \left( \overline{F^+_R}  \eta^+   
- \frac{1}{\sqrt{2}}
 \overline{F^0_R} \eta^0 \right) \hat \nu_{L}
\end{align}
and
\begin{align}
\mathcal{L} \supset y_2 \overline  F_L \,\eta\, \nu_R \supset y_2 \left( \overline{F^+_L}  \eta^+  
+ \overline{F^0_L} \eta^0  \right) \hat \nu_R.    
\end{align}

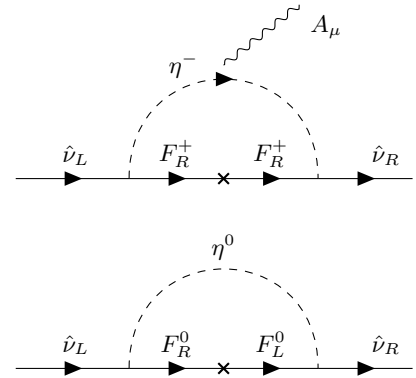
\begin{figure}[b!]
\centering
\begin{subfigure}{0.45\textwidth}
\centering
\begin{tikzpicture}
  \begin{feynman}
    \vertex (a1) at (0,0);                          
    \vertex [right=1.5cm of a1] (aPsi);             
    \vertex [right=2.5cm of aPsi] (aF);             
    \vertex [right=1.25cm of aF] (anuR);            

    \coordinate (amidferm) at ($(aPsi)!0.5!(aF)$);   

    \vertex (aSloopTop) at ($(aPsi)!0.5!(aF)+(0,1.5)$);
    \vertex (aboson) at ($(aSloopTop)+(1,0.8)$);    

    \draw[thick] ([xshift=-2pt,yshift=-2pt]amidferm) -- ([xshift=2pt,yshift=2pt]amidferm);
    \draw[thick] ([xshift=-2pt,yshift=2pt]amidferm) -- ([xshift=2pt,yshift=-2pt]amidferm);

    \node[above=2pt of aPsi, xshift=-20pt] {\(\hat\nu_L\)};
    \node[above=2pt of anuR, xshift=-10pt] {\(\hat\nu_R\)};
    \node[above=-8pt of aSloopTop, xshift=-15pt] {\(\eta^-\)};
    \node[below right=1pt and 1pt of aboson] {\(A_\mu\)};
    \node[above=1pt of $(aPsi)!0.5!(amidferm)$] {\(F^+_R\)};
    \node[above=1pt of $(amidferm)!0.5!(aF)$] {\(F^+_R\)};

    \diagram* {
      (a1) -- [fermion] (aPsi),
      (aPsi) -- [fermion] (amidferm),
      (amidferm) -- [fermion] (aF) -- [fermion] (anuR),
      (aPsi) -- [half left, charged scalar, dashed, looseness=1.8] (aF),
      (aSloopTop) -- [boson] (aboson),
    };
\end{feynman}
\end{tikzpicture}
\end{subfigure}

\vspace{0.5cm}

\begin{subfigure}{0.45\textwidth}
\centering
\begin{tikzpicture}
  \begin{feynman}

    \vertex (b1) at ($(anuR)+(2.5,0)$);                           
    \vertex [right=1.5cm of b1] (bPsi);             
    \vertex [right=2.5cm of bPsi] (bF);             
    \vertex [right=1.25cm of bF] (bnuR);            

    \coordinate (bmidferm) at ($(bPsi)!0.5!(bF)$);   

    \vertex (bSloopTop) at ($(bPsi)!0.5!(bF)+(0,1.5)$);

    \draw[thick] ([xshift=-2pt,yshift=-2pt]bmidferm) -- ([xshift=2pt,yshift=2pt]bmidferm);
    \draw[thick] ([xshift=-2pt,yshift=2pt]bmidferm) -- ([xshift=2pt,yshift=-2pt]bmidferm);

    \node[above=2pt of bPsi, xshift=-20pt] {\(\hat\nu_L\)};
    \node[above=2pt of bnuR, xshift=-10pt] {\(\hat\nu_R\)};
    \node[above=-5pt of bSloopTop] {\(\eta^0\)};
    \node[above=1pt of $(bPsi)!0.5!(bmidferm)$] {\(F^0_R\)};
    \node[above=1pt of $(bmidferm)!0.5!(bF)$] {\(F^0_L\)};

    \diagram* {
      (b1) -- [fermion] (bPsi),
      (bPsi) -- [fermion] (bmidferm),
      (bmidferm) -- [fermion] (bF) -- [fermion] (bnuR),
      (bPsi) -- [half left, scalar, dashed, looseness=1.8] (bF),
    };

  \end{feynman}
\end{tikzpicture}
\end{subfigure}
\caption{Model-II. Top panel: Leading contribution to NMM; once the photon leg is removed it also contributes to neutrino mass.  Bottom   panel: contribution to  neutrino mass via neutral BSM states.}
\label{fig:900}
\end{figure}

Utilizing these interactions, the leading contribution to the NMM arises from the diagram shown in the top panel of Fig.~\ref{fig:900}. Upon removing the photon leg, this diagram generates a neutrino mass proportional to $m_\nu^{\mathrm{left}} \propto -y_{\Psi} y_1 y_2$. On the other hand, an additional contribution to the neutrino mass arises from the diagram shown in the  bottom panel of Fig.~\ref{fig:900}, yielding $m_\nu^{\mathrm{right}} \propto y_{\Psi} y_1 y_2$, with the opposite sign. Therefore, a cancellation due to the triplet mechanism is expected in this scenario, provided that the BSM charged states ($F^\pm, \eta^\pm$) and their corresponding neutral states ($F^0, \eta^0$) are degenerate in mass. However, EW symmetry breaking inevitably induces mass splittings~\cite{Cirelli:2005uq} among the components of multiplets transforming nontrivially under $SU(2)_L$. Once these corrections are taken into account, the exact cancellation is destabilized, leading to unacceptably large neutrino masses.
In the next section, we provide a concrete example with colored states, where these essential EW corrections are computed explicitly.

Therefore, without invoking fine-tuning—such as cancellations between these loop-induced contributions and the tree-level mass—an enhanced NMM cannot be achieved in a phenomenologically viable scenario.

\section{Extended Model with Colored States}\label{sec:03}
The mixing between the SM Higgs and the BSM neutral scalars can also be forbidden without introducing additional symmetries. A simple realization of this occurs when the new states carry color. In this section, we explore such a scenario—consisting of colored states\footnote{
Scalar leptoquarks employed to generate the neutrino magnetic moment in a different context have been discussed, for example, in Ref.~\cite{Brdar:2020quo}.
}—which we label as Model-III.

In addition to $\Psi_{L,R} \sim (1,3,0;1)$, we introduce a colored fermion $F \sim (\overline{3},3,1/3;1)$ and a scalar $S \sim (\overline{3},3,1/3;0)$, which share the same SM gauge quantum numbers but differ in their lepton number assignments. For concreteness, we write these fields explicitly in components,
\begin{align}
&F= \begin{pmatrix}
F^{+1/3}/\sqrt{2}  & F^{+4/3}
\\
F^{-2/3}  & - F^{+1/3}/\sqrt{2}
\end{pmatrix},    
\\ &
S= \begin{pmatrix}
S^{+1/3}/\sqrt{2}  & S^{+4/3}
\\
S^{-2/3}  & - S^{+1/3}/\sqrt{2}
\end{pmatrix}. 
\end{align}
It is worth noting that since the colored scalar does not carry any lepton number—which is conserved in our scenario—a lepton-quark coupling cannot be written down. However, it can have a di-quark coupling only; therefore, proton decay is not a problem.

The presence of these states permits the construction of the following relevant terms:
\begin{align}
\mathcal{L}_Y &\supset  m_F \;\mathrm{tr}\left(\overline F_L F_R\right)  +  y_T \;\mathrm{tr}\left(\overline \Psi_L S^* F_R\right)  \nonumber\\ &+
y_\delta \;\mathrm{tr}\left(\overline F_L S\right) \nu_R  +h.c.\;.
\end{align}
The first term provides equal masses ($=m_F$) to the $F^{4/3}, F^{1/3}, F^{-2/3}$ state. 
Now, building on the discussion in the previous sections and working in the mass basis, we can write
\begin{align}
y_T &\;\mathrm{tr}\left(\overline \Psi_L S^* F_R\right)    \supset 
\nonumber \\& 
\frac{y_T}{\sqrt{2}} \overline{\hat \Psi^-}_L   \left(  S^{-4/3} F_R^{+1/3} - S^{-1/3} F_R^{-2/3} \right) 
\nonumber \\&  
+\frac{y_T}{\sqrt{2}} \overline{\hat \Psi^+}_L   \left(  S^{-1/3} F_R^{+4/3} - S^{+2/3} F_R^{+1/3} \right) 
\nonumber \\&   
+\frac{y_\Psi\, y_T\, v_H}{2m_\Psi} \overline{\hat\nu}_L \left( F^{+4/3}_R  S^{-4/3} -  F^{-2/3}_R S^{+2/3} \right), \label{eq:color-1}
\end{align}
and
\begin{align}
y_\delta &\;\mathrm{tr}\left(\overline F_L S\right) \nu_R   \supset 
\frac{y_\Psi\, y_D \,y_\delta\, v_H^2}{\sqrt{2}m^2_\Psi} \bigg( \overline{F^{-2/3}}_L  S^{-2/3}
\nonumber \\&
+ \overline{F^{+1/3}}_L  S^{+1/3} 
+ \overline{F^{+4/3}}_L  S^{+4/3} \bigg) \hat\Psi^0_R
\nonumber \\&+ 
y_\delta \bigg( \overline{F^{-2/3}}_L  S^{-2/3} 
+ \overline{F^{+1/3}}_L  S^{+1/3} + \overline{F^{+4/3}}_L  S^{+4/3}  \bigg) \hat\nu_R .  \label{eq:color-2}
\end{align}

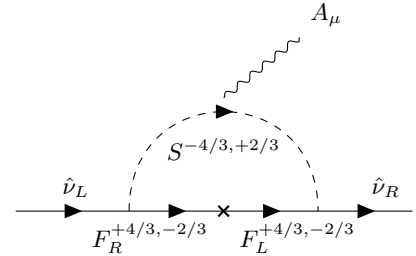
\begin{figure}[th!]
\centering
\begin{tikzpicture}
  \begin{feynman}

    \vertex (v1) at (0,0);                              
    \vertex [right=1.5cm of v1] (Psi);                  
    \vertex [right=2.5cm of Psi] (F);                   
    \vertex [right=1.25cm of F] (nuR);                  

    \coordinate (midferm) at ($(Psi)!0.5!(F)$);         

    \vertex (SloopTop) at ($(Psi)!0.5!(F)+(0,1.5)$);    
    \vertex (boson) at ($(SloopTop)+(1,0.8)$);          

    \draw[thick] ([xshift=-2pt,yshift=-2pt]midferm) -- ([xshift=2pt,yshift=2pt]midferm);
    \draw[thick] ([xshift=-2pt,yshift=2pt]midferm) -- ([xshift=2pt,yshift=-2pt]midferm);

    \node[above=2pt of Psi, xshift=-20pt] {\(\hat\nu_L\)};
    \node[above=2pt of nuR, xshift=-10pt] {\(\hat\nu_R\)};
    \node[below=12pt of SloopTop] {\(S^{-4/3,+2/3}\)};
    \node[above right=1pt and 1pt of boson] {\(A_\mu\)};

    \node[below=1pt of $(Psi)!0.5!(midferm)$, xshift=-10pt] {\(F_R^{+4/3,-2/3}\)};
    \node[below=1pt of $(midferm)!0.5!(F)$, xshift=10pt] {\(F_L^{+4/3,-2/3}\)};

    \diagram* {
      (v1) -- [fermion] (Psi),
      (Psi) -- [fermion] (midferm),
      (midferm) -- [fermion] (F) -- [fermion] (nuR),
      (Psi) -- [half left, charged scalar, dashed, looseness=1.8] (F),
      (SloopTop) -- [boson] (boson),
    };

  \end{feynman}
\end{tikzpicture}
\caption{Model-III. Leading contributions to the NMM in the scenario with colored states. Removing the external photon leg yields one-loop neutrino mass contributions from two separate diagrams: one involving the BSM colored states $F^{\pm 4/3}, S^{\pm 4/3}$, and the other involving $F^{\pm 2/3}, S^{\pm 2/3}$.} \label{fig:05}
\end{figure}

These interactions can, in principle, generate a large NMM, as illustrated in Fig.~\ref{fig:05}. Upon removing the photon leg, the corresponding diagrams contribute to the neutrino mass, which depends on the couplings given in Eqs.~\eqref{eq:color-1}–\eqref{eq:color-2}. These equations show explicitly that the vertices $\overline{\hat\nu}_L F^{+4/3}_R S^{-4/3}$ and $\overline{\hat\nu}_L F^{-2/3}_R S^{+2/3}$ have opposite signs, whereas the vertices $\overline{F^{-2/3}}_L S^{-2/3} \hat\nu_R$ and $\overline{F^{+4/3}}_L S^{+4/3} \hat\nu_R$ share the same sign, resulting in an exact cancellation. In contrast, the NMM diagram in Fig.~\ref{fig:05}, which includes the photon leg, incorporates the electromagnetic charges ($4/3$ and $-2/3$), leading to nonvanishing and potentially large contributions.

Note, however, that EW symmetry breaking inherently lifts the mass degeneracy among the different components of the scalar multiplet, already at tree level. Even if this tree-level mass splitting is tuned to zero by an appropriate choice of parameters, radiative corrections inevitably reintroduce mass splittings and spoil the cancellation. We first analyze the tree-level mass splitting of the scalar components and then explicitly examine the unavoidable radiative corrections and their impact on the NMM.

In this setup, the NMM and neutrino masses are  found to be~\cite{Lavoura:2003xp}  
\begin{align}
\mu_\nu  &=  \frac{N_c y_\Psi y_T y_\delta}{12\pi^2} \frac{v_H}{m_\Psi}
\bigg\{
m_{F_{4/3}} m^2_{S_{4/3}}  
\nonumber \\&
\frac{  2(m^2_{S_{4/3}}-m^2_{F_{4/3}}) + (m^2_{S_{4/3}}+m^2_{F_{4/3}})  \log \frac{m^2_{F_{4/3}}}{m^2_{S_{4/3}}} }{(m^2_{S_{4/3}}-m^2_{F_{4/3}})^3} 
\nonumber\\&
+
m_{F_{2/3}} m^2_{S_{2/3}}  
\nonumber \\&
\frac{  2(m^2_{S_{2/3}}-m^2_{F_{2/3}}) + (m^2_{S_{2/3}}+m^2_{F_{2/3}})  \log \frac{m^2_{F_{2/3}}}{m^2_{S_{2/3}}} }{(m^2_{S_{2/3}}-m^2_{F_{2/3}})^3} 
\bigg\},
\end{align}
and
\begin{align}
m_\nu^\mathrm{1-loop}=     \frac{N_c y_\Psi y_T y_\delta}{32\pi^2} & \frac{v_H}{m_\Psi}    \bigg\{ m_{F_{4/3}}
\frac{m^2_{S_{4/3}} \log \frac{m^2_{S_{4/3}} }{m^2_{F_{4/3}}} }{m^2_{S_{4/3}}-m^2_{F_{4/3}}}
\nonumber \\&
-
m_{F_{2/3}}
\frac{m^2_{S_{2/3}} \log \frac{m^2_{S_{2/3}} }{m^2_{F_{2/3}}} }{m^2_{S_{2/3}}-m^2_{F_{2/3}}}
\bigg\}.
\end{align}
In the limit where $m_{F_{4/3}}=m_{F_{2/3}}$ and $m_{S_{4/3}}=m_{S_{2/3}}$, we have $m_\nu^\mathrm{1-loop}=0$, while the NMM remains non-zero, and can be potentially large. 

As mentioned above, at tree level we have $m_{F_{4/3}}=m_{F_{2/3}}$; however, the corresponding degeneracy in the scalar sector generally does not hold. To see this, let us consider the  relevant scalar potential 
    \begin{align}
        V\supset & \; m^2_S \;\mathrm{tr}\left(S^\dagger S\right) + \lambda H^\dagger H \;\mathrm{tr}\left(S^\dagger S\right) 
        \nonumber\\&
        +i\epsilon^{abc} \lambda^\prime \left( H^\dagger \sigma_a H\right) S^{b\dagger} S^c \;.
    \end{align}
In the last term, $\sigma_a$ are the Pauli matrices and the field, $S$,  is written in the triplet representation. Once the EW symmetry is broken, we find 
\begin{align}
&m^2_{S_{4/3}}=m^2_{S_{1/3}} + \lambda^\prime v_H^2,
\quad
&m^2_{S_{2/3}}=m^2_{S_{1/3}} - \lambda^\prime v_H^2.
\end{align}

\begin{figure}[t!]
\centering
\includegraphics[width=0.4\textwidth]{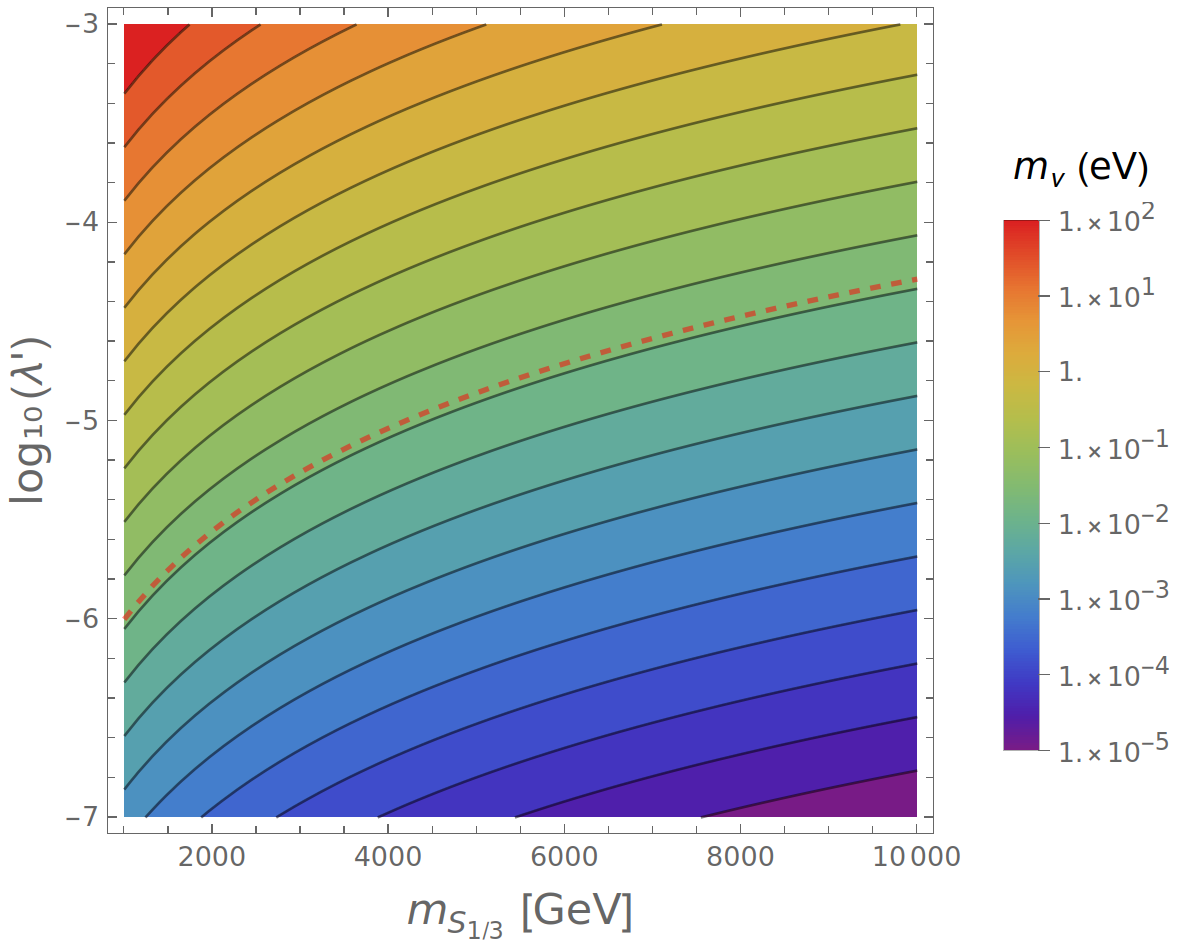}
\includegraphics[width=0.4\textwidth]{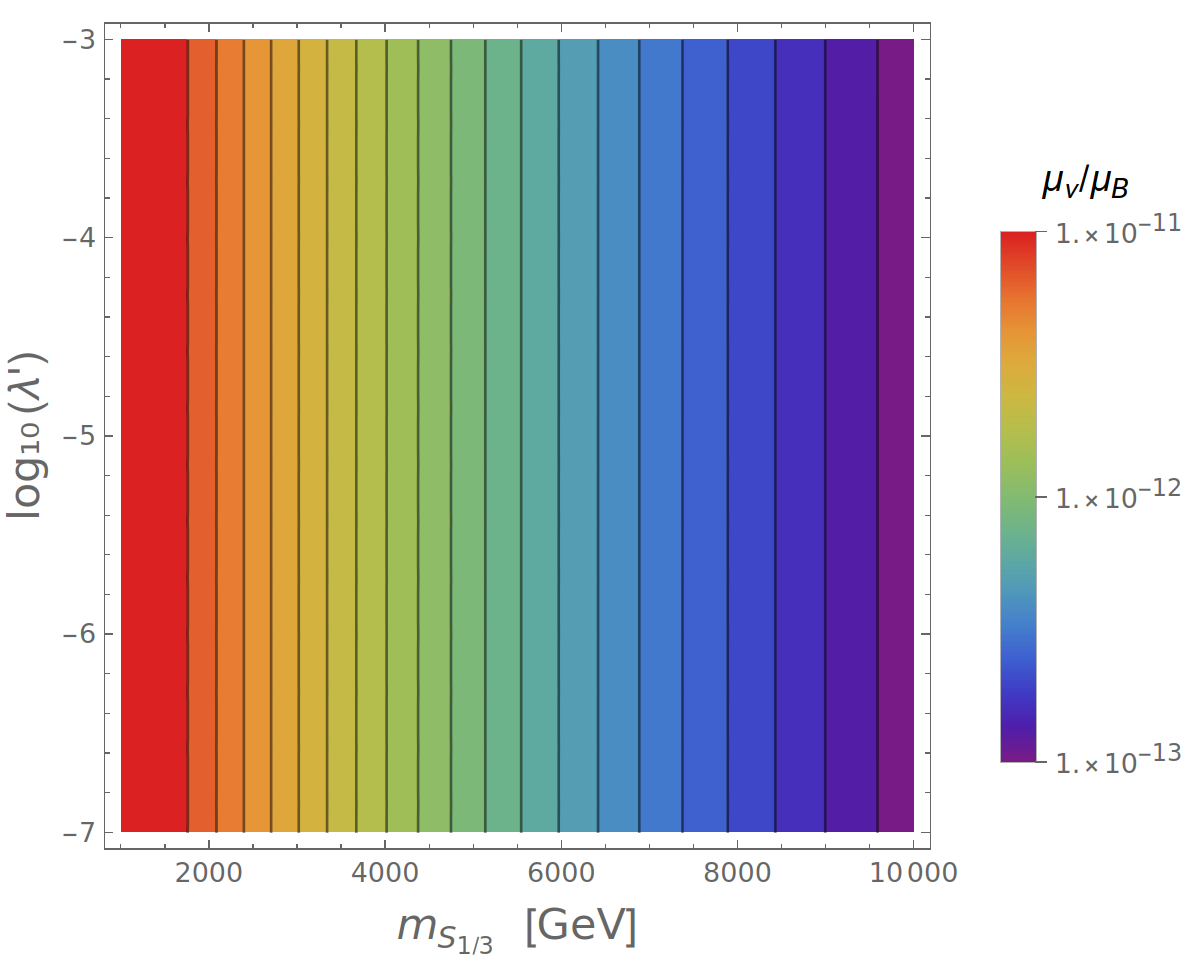}
\caption{Model-III. Top panel: Demonstrating that a value  $\lambda^\prime\sim (0.1-1)\times 10^{-5}\sim y_e$ is required to reproduce the observed neutrino mass scale for BSM colored states with masses in the multi-TeV range. The red dashed line corresponds to the observed neutrino mass scale $m_\nu = 0.05$ eV.  Bottom panel:  To a very good approximation, NMM is independent of the tree level mass splitting that as long as $\lambda^\prime \ll 1$. NMM in the observable range is obtained with  multi-TeV BSM states.  Radiative corrections to the BSM masses are not included in these plots for illustration. } \label{fig:LQ}
\end{figure}

The dimensionless parameter $\lambda'$ is constrained by EW precision measurements. The aforementioned mass non-degeneracy impacts the oblique parameters, and therefore, the mass splittings are typically required to satisfy $\Delta m_S \lesssim 50~\mathrm{GeV}$~\cite{Dorsner:2016wpm,Chowdhury:2022dps}. For our scenario, any  mass splitting needs to be extremely small not to overshoot the neutrino masses; hence, the oblique parameters do not provide any meaningful constraints. As a consequence of this non-degeneracy arising from  $\lambda^\prime\neq 0$, we obtain $m_\nu^\mathrm{1-loop}\neq 0$.  However, as  demonstrated in Fig.~\ref{fig:LQ} (top panel), for BSM states in the multi TeV range, neutrino masses on the order of $m_\nu^\mathrm{1-loop}\sim 0.05$ eV can be reproduced if  $\lambda^\prime\sim (0.1-1)\times 10^{-5}\sim \mathcal{O}(y_e)$. Meanwhile, the NMM, which is practically independent of $\lambda^\prime$ (see bottom of Fig.~\ref{fig:LQ}), can remain within the observable range of future experiments.

\begin{figure}[t!]
\centering
\includegraphics[width=0.4\textwidth]{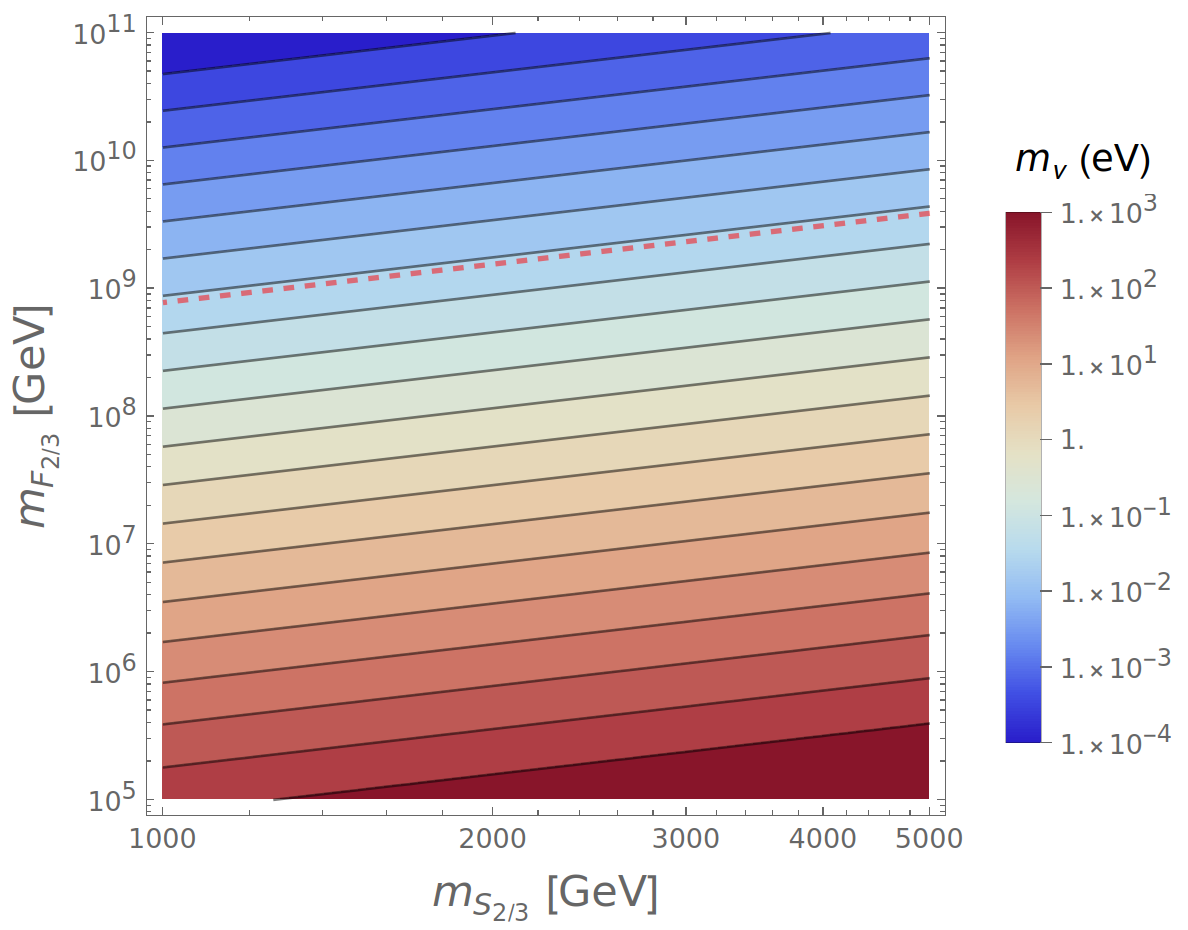}
\includegraphics[width=0.4\textwidth]{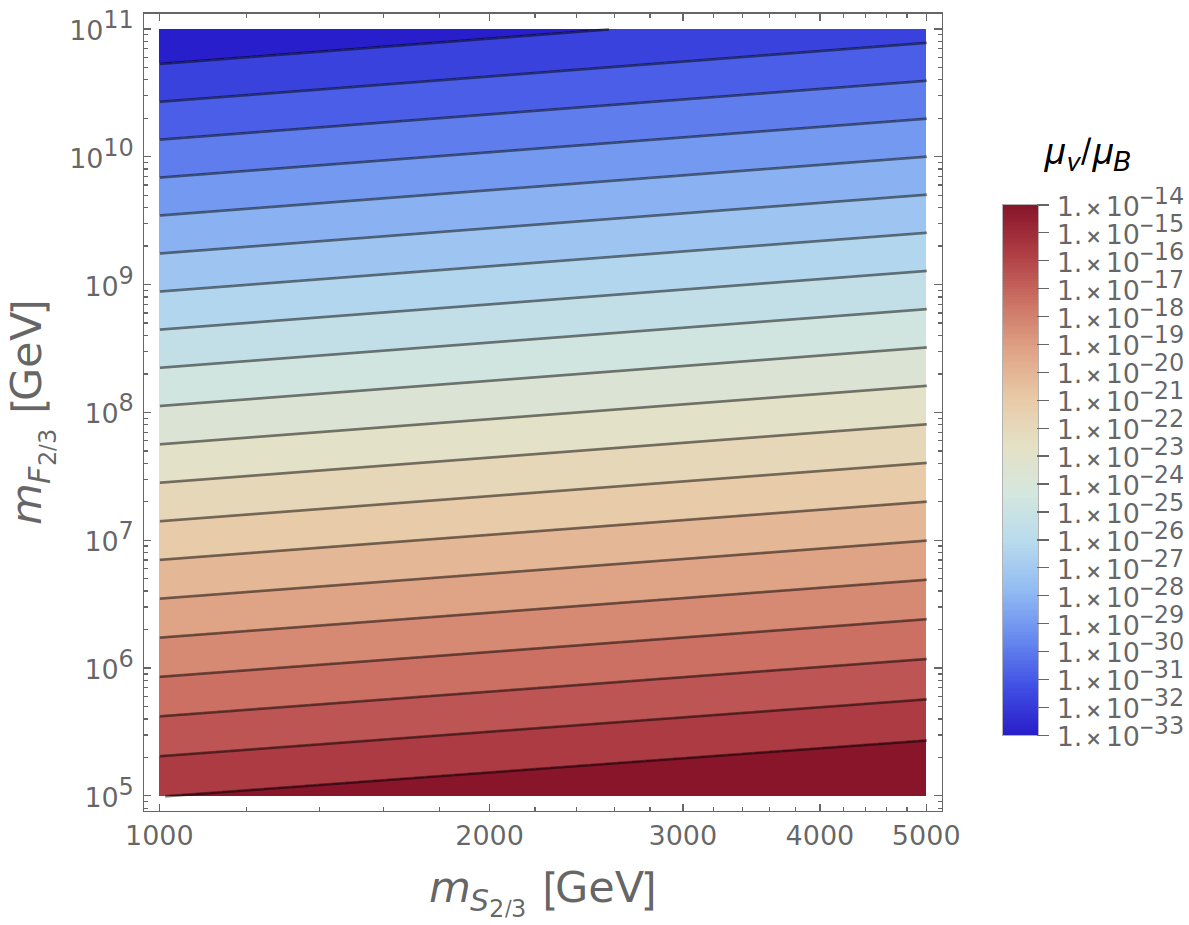}
\caption{Model-III. Top panel: EW symmetry breaking induces mass splittings that completely destabilize the exact cancellation in the one-loop neutrino mass diagram, requiring the new physics scale to be extremely large in order not to exceed the observed neutrino mass scale.  The red dashed line corresponds to the observed neutrino mass scale $m_\nu = 0.05$ eV. Bottom panel: Such a high new physics scale effectively drives the NMM to zero, placing it far beyond any observable range. } \label{fig:LQ:loop}
\end{figure}

However, this mechanism can be destabilized, again via thr EW symmetry breaking,  which introduces radiative corrections that spoils the mass degeneracy among the fermionic and scalar components. In particular, once these quantum corrections are included, the one-loop neutrino mass $m_\nu^\mathrm{1-loop}$ can  become larger than the observed scale by several orders of magnitude. To simplify the analysis, we assume that all components share the same mass at tree level. To quantify this, let us first write down the mass splitting  induced by loops of SM gauge bosons between two components of $F, S$~\cite{Cirelli:2005uq}
\begin{align}
m^2_{F_{4/3}}&=m^2_{F_{2/3}} + \frac{\alpha_2 s_W^2 M_Z}{6\pi} \bigg\{  
-2r_F+2r^3_F \log(r_F)
\nonumber \\ & \hspace{30pt}
+ (-4+r^2_F)^{1/2} (2+r^2_F) \log(A_F)
\bigg\},
\\
m^2_{S_{4/3}}&=m^2_{S_{2/3}} + \frac{\alpha_2 s_W^2 M_Z}{12\pi} \bigg\{  
-2r^3_S \log(r_S)
\nonumber \\ & \hspace{50pt}
- (-4+r^2_S)^{3/2} \log(A_S)
\bigg\},
\end{align}
where we have defined $r_X=M_Z/m_X$ and $A_X=-1+\frac{1}{2} r^2_X   - \frac{1}{2} r_X  (-4+r^2_X)^{1/2}$ (with $X=F$ or $S$). In the limit $m_X \gg M_Z$, the one-loop corrections lead to $\Delta m_X\simeq 470$ MeV for both fermions and scalars. Although this correction to the order may seem small for $m_{F,S} \gtrsim$ TeV, the neutrino mass, due to its tininess, is in fact sensitive to it. As can be seen from Fig.~\ref{fig:LQ:loop} (top panel), only in the large mass limit of the fermion, namely, $m_{2/3}\sim 10^9$ GeV, correct neutrino mass scale is reproduced.  However, in this  limit, the NMM lies far below current experimental sensitivities; see the bottom panel in Fig.~\ref{fig:LQ:loop}.   

In summary, unless fine-tuning is invoked—for example, through a cancellation between the large one-loop contribution and a tree-level contribution—a large NMM within the observable range cannot be achieved within this mechanism.

\vspace{10pt}
\textbf{Key Findings}-- 
\begin{itemize}
\item In the minimal model—extending the SM by a Dirac fermion pair and a real scalar, both transforming as weak triplets with zero hypercharge—there exists an additional one-loop contribution to the neutrino mass, not considered in Ref.~\cite{Pages:2025odc}, arising from the propagation of neutral scalars. In this setup, the neutrino magnetic moment   is decoupled from the neutrino mass. Achieving a NMM close to current experimental bounds then requires unsuppressed Yukawa couplings.     However, in this case, the corresponding mixing angle between the SM Higgs and the BSM scalar must be smaller than $\mathcal{O}(10^{-11})$ to avoid exceeding neutrino mass constraints. This immediately leads to severe fine-tuning of the model parameters, in particular among the cubic coupling and the product of the vacuum expectation value and a quartic coupling.

\item More generally, in any model where the additional scalar field—whether the weak triplet of the minimal model or any other representation containing a neutral component—mixes with the SM Higgs, a similar fine-tuning problem is unavoidable.

\item On the other hand, if the BSM scalar is forbidden from mixing with the SM Higgs—for example, by imposing a discrete symmetry or by carrying additional charges such as color—the above source of fine-tuning can be avoided. However, apart from the minimal model, all extended scenarios require BSM states with nonzero hypercharge and therefore generically suffer from another issue: EW symmetry breaking induces mass splittings among components with different electric charges at the loop level, thereby destabilizing the cancellation between neutrino mass contributions.
\end{itemize}

Therefore, our careful analysis shows that the decoupling of the neutrino magnetic moment from the neutrino mass, is not realized in general within this framework unless finely tuned cancellations are invoked.

\section{Conclusions}\label{sec:04}
In conclusion, we have examined the framework of the weak triplet mechanism, which aims to reconcile large neutrino magnetic moments—potentially within reach of upcoming experiments—with naturally small Dirac neutrino masses. While this setup introduces new interactions through the mixing of light neutrinos with beyond the Standard Model neutral fermions, thereby allowing contributions to the neutrino magnetic moment that are not directly proportional to the neutrino mass, we find that this decoupling is only apparent.

In the minimal realization, the neutrino magnetic moment can in principle be decoupled from the neutrino mass; however, achieving a sizable enhancement requires significant fine-tuning to suppress the mixing between the SM Higgs and the neutral BSM scalar, so as not to exceed neutrino mass constraints. In this minimal version of the model, even in the limit where the trilinear coupling of the triplet scalar to the SM Higgs is taken to be very small, i.e., $\mu \ll 1\,\text{GeV}$ as in Ref.~\cite{Pages:2025odc}, an observable NMM can be achieved only in a very special part of the parameter space, requiring the product of the corresponding triplet VEV and the quartic coupling between the triplet and the SM Higgs to be small; this smallness is not guaranteed by the same symmetry argument that justifies a small $\mu$. Alternatively, assuming a positive mass term (\(-\mu^2_S > 0\)) in Eq.~\ref{eq:fullpot} prevents \(S\) from acquiring a vacuum expectation value during spontaneous symmetry breaking. Under this condition, Eq.~\ref{eq:EWcondition} isolates the \(\mu \) term, meaning the primary constraint is fulfilled only as \(\mu \to 0\).

In contrast, in all extended scenarios, the decoupling is no longer realized. In particular, although the corresponding neutrino mass contributions vanish in the limit of unbroken electroweak symmetry (as it should!), electroweak symmetry breaking inevitably reintroduces a connection between the neutrino mass and magnetic moment. As a result, any significant enhancement of the neutrino magnetic moment within this framework requires severe fine-tuning of the neutrino mass. 

Our results therefore indicate that, within the weak triplet mechanism, achieving an observable neutrino magnetic moment is generically incompatible with maintaining naturally small Dirac neutrino masses. This underscores the persistent challenge of the magnetic moment–mass problem and motivates the exploration of alternative frameworks beyond those considered here.

\section*{Acknowledgments}
S.S.\ would like to thank Kaladi Babu for valuable discussions. We also thank Julie Pagès and Anil Thapa for their comments on our manuscript.   S.F.\ and S.S.\ acknowledge the financial support
from the Slovenian Research Agency (research core funding No.\ P1-0035 and N1-0321).

\bibliographystyle{style}
\bibliography{reference}
\end{document}